# Quantum Mechanics: Incomplete and Non Local Theory

M. Cattani

Institute of Physics, University of S. Paulo, C.P. 66318, CEP 05315–970 S. Paulo, S. P. Brazil. E–mail: mcattani@if.usp.br

**Abstract.**
We will show for undergraduate and graduate students of physics that Quantum Mechanics is an incomplete and non–local theory. The problem of non–locality is discussed by analyzing the Bell's theorem where are considered correlations between measurements results performed on physical systems that are far apart, but that interacted in the past. The experimental violations of Bell's theorem show a very general result that quantum phenomena are nonlocal and that, inevitably, Quantum Mechanics is non–local.
*Key words. Quantum Mechanics. Bell theorem. incomplete and non–local theory.*

## I) Introduction.

Our goal is to show in a simplest way for undergraduate and graduate students in physics that Quantum Mechanics (QM) is an *incomplete and non-local theory*. To do this we define in Classical Mechanics (CM) and Quantum Mechanics (QM) the meanings of *action at a distance, locality, causality, determinism and theory complete and incomplete*. We do this analysis because these concepts are seen very briefly in the basic undergraduate physics course. Our analysis will be done primarily keeping in mind that science is concerned only with observable things. Our conclusions are obtained by comparing the theoretical predictions with results of measures actually taken or realizable in physical systems. We use purely phenomenological arguments, avoid talking about realism, idealism, hidden variables, etc.. In other words, electrons, photons, measuring instruments, etc., are real things and the variables used are those normally adopted in the textbooks for undergraduate and graduate students in physics. We define Quantum Mechanics (QM) the theory constructed using the *usual variables* found in basic textbooks for graduate and postgraduate level. We do this to distinguish it from a quantum theory that can be constructed using *hidden variables* (see Section 6) that can be different from the *usual* ones. Following the same procedure adopted in our previous didactical articles, we will quote the lowest possible number of books and articles on the subject. In Section 1 we see that the non-relativistic CM is a *non−local, causal, deterministic and complete theory* and that the relativistic CM is *local, causal, deterministic and complete.* In section 2 show the meaning of causality and determinism in QM and show that it is an *incomplete theory*. In Sections 3 we analyze the correlations between measurements in CM and in QM. In Section 4 we deduce Bell's theorem and examine the problem of non−locality of quantum theory based on experimental violations of this theorem observed in quantum correlations. In Section 4 we conclude that *quantum theory is non−local* and that *QM is incomplete and non−local*. In Section 5 we make brief comments on *hidden variables theories* and present the final conclusions.

## 1) Classical Mechanics.

1.a) *Non-Relativistic Classical Mechanics. Causality*.

In the basic courses of mechanics and electromagnetism[1-4] "ab initio" it is assumed that the gravitational and electromagnetic forces between the bodies (with charge q and mass m) are instantaneous interactions that are also called non−local interactions. Analyzing the causality problem we must be aware that we are venturing into controversial areas where it is difficult to proceed entirely free from bias. A broad and thorough discussion on the topic can be found, for example, the book by Lindsay and Margenau.[5] We will avoid philosophical discussions and subtle problems in physics such as microscopic reversibility and irreversibility and similar ones. These are covered in the book cited above.[5]

Classical Mechanics (CM) has been developed continuously since the time of Newton and applied to increasingly larger number of dynamical systems, including the electromagnetic field interacting with matter. Let us use the trajectory **r**(t) of a particle to define causality in CM. Suppose the particle is subjected to a well known conservative force **F**(**r**). Thus, causality can be defined (or characterized) assuming:[5,6] "The differential equations describing the motion are not explicit functions of time." Thus, to determine the trajectory **r**(t) we must integrate the differential equation (Newton's second law)

$$m \, d^2\mathbf{r}/dt^2 = \mathbf{F}(\mathbf{r}) \qquad (1.1),$$

knowing the (initial) position and velocity **r**($t_o$) and **v**($t_o$), respectively, at a given moment (initial) t = $t_o$. With this procedure the position **r**(t) of the particle is mathematically determined at any time t, past or future, i.e. for t ≤ to or t ≥ to. Under these conditions, we say that the CM is *non-local, causal and deterministic.*

Without changing the above findings (1.1) describes a system consisting of two particles simply replacing m by the reduced mass µ of the bodies and considering **F**(**r**) as the force and **r**(t) the relative distance between the bodies,[7] respectively. In the conservative case the total energy E contained in a binary system of particles is given by E = µ**v**$^2$/2 + V(**r**), where V(**r**) is the potential energy of the system and **F**(**r**) = − grad V(**r**). It is assumed that the energy spectrum E is continuous.

From above exposed, it is assumed that causality applies only to systems that remain undisturbed. But we must remember that science is concerned only with observable things and to observe an object it must interact with an external agent. Thus, an act of observation is necessarily accompanied by a perturbation of observed object.[6] Based on nearly a century of existence of QM and extremely elaborate experimental techniques that were developed and used to obtain a multitude of experimental results we define an object as *big* when the disturbance created by the observation is negligible and as *small* when the disturbance can not be neglected.[6] Thus, if an object can be observed so that the inevitable disruption created by observation is negligible we say that it is big and that CM can be applied to him. This implies that the observed trajectory of a particle in the CM of a particle is exactly described by the function **r**(t). This occurs, for instance, for macroscopic bodies such as planets, tennis balls, etc. .... Knowing **r**(t) we can exactly determine (predict) all physical properties of the particle, **p** = m**v**, m**v**$^2$/2, **L** = **r** x **p**, etc., involved in the mechanical phenomena. So we say that the CM is a *complete theory*. It is important to note that microscopic charged particles such as protons, electrons, positrons, mesons, α particles,…, generate dots and dashes in bubble chambers and emulsions.[8] These dots and dashes are displayed along a filament with a "trajectory" that obeys (1.1). As the filament diameter ~ 10$^3$ larger than the diameter of the detected particles the function **r**(t) do not really describe the particle trajectory.

1.b) *Relativistic Classical Mechanics.*

In the Special Relativity (SR)[1–4] course we learn that "nothing (object or signal) can propagate with a velocity greater than the speed of light c in vacuum." Thus, according to the SR the interactions between objects are *local* : "actions that occur at a given point P of space does not have any instantaneous effect in another distant point of P".[9] With the advent of the SR it was necessary to reformulate all CM laws writing them in a covariant form and adopting, for example, the lagrangian and hamiltonian formalisms.[7,9] For particles submitted to velocity independent conservative forces the relativistic lagrangian L is given by L = −mc$^2$γ

−V(**r**), where V(**r**) is a scalar potential and $\gamma^{-1} = [1-(v/c)^2]/2$. The hamiltonian H is given by $H = T + V = E = mc^2/\gamma + V(\mathbf{r})$. The lagrangian for a single particle with charge q in an electromagnetic field is given by $L = -mc^2\gamma - q\varphi + (q/c)v$ where $A_\mu = (\varphi, \mathbf{A})$ is the 4−vector electromagnetic potential. In the case of conservative systems L and H functions are functions only of **r** and **v** and not depending explicitly on the time t. Using the lagrangian and hamiltonian formalisms[7,9] the particles trajectories **r**(t) can be determined mathematically, for any time t, solving differential equations by knowing the initial conditions $\mathbf{r}(t_o)$ and $\mathbf{v}(t_o)$. So, the relativistic CM is *local, causal, deterministic* and *complete*. It can be only applied to large objects.

*Comments on objects and signals.*

An *object* would be a particle with a non zero inertial mass m: its speed could never reach or exceed the speed of light. A *signal* would be an "entity" without inertial mass. As we know, from the SR there is no "entity" that can propagate faster than c. Only an electromagnetic wave (EW) propagates with the speed c. It can have only one frequency or be formed by a "package" of frequencies. In both cases, the EW is an entity that would transmit information. In Section 5 we comment on possible supra−luminous signals that could be able to transfer information between two quantum correlated particles (*"entangled states of pairs of particles"*).

1c) *Conclusions.*

The non−relativistic CM is a *causal, deterministic, non−local and complete theory*. The relativistic CM is a *causal, deterministic, local and complete theory*.

## 2).Quantum Mechanics.

In the late 19th century there were several phenomena such as black body spectrum, photo-electric effect, Compton effect and atomic spectra of discrete rays that could not be explained with classical physics.[1-4]. It was necessary to introduce news and revolutionary concepts to explain them. In 1901 was created by Max Planck[1−4] a new physics called quantum physics. Postulating that the energies of particles could take discrete values he explained the blackbody spectrum. With this hypothesis Einstein defined photons and explained the photoelectric effect; using photons Compton effect was explained. De Broglie postulating the wave−particle duality explained the wave behavior of particles. Leaning on Planck, Einstein and de Broglie, Bohr explained the discrete ray spectrum of the hydrogen.[1-4] For the time those ideas that seemed so absurd were proposed because classical physics could not explain the aforementioned phenomena occurring on a microscopic scale, i.e., those involving electrons, protons, photons, atoms and molecules. Thus physics was divided roughly into two parts: the classical physics that describe the macroscopic world and quantum physics that should describe the phenomena in atomic scale, i.e., those occurring in the microscopic world. Thus, the concept of *big* and *small* left to be simply a relative concept [6] to have an absolute meaning. With the advent of quantum physics, we define an object as *big* when the disturbance generated by the observation may be neglected and small when the disturbance can not be neglected. At least until today we can say that atomic particles and sub-atomic are *small*. Only in 1925–1926 it was established a microscopic theory called Quantum Mechanics (QM).[8-13] Heisenberg proposed a matrix mechanics and Schrödinger a wave mechanics, both non−relativistic. The matrix formulation is useful for studying problems involving harmonic oscillator and angular momentum, but for others it is a bit difficult to use. "En passant", we would stress that there are essentially nine different formulations of non−relativistic quantum equations.[14] In 1927−1928 two quantum relativistic wave equations[10−11] were constructed to particles with spin 0 and spin ½. The first (spin 0) was proposed by Klein, Fock and Gordon and the second (spin ½) by Dirac. The strong wave character of particles in atomic and subatomic dimensions obliged the physicists to abandon the description of the microscopic phenomena based on the particles trajectories calculations. Instead of saying that a particle is at a point **r**(t) they started to say that there is a probability to find it in a volume element $d^3$**r** around **r**(t). The wave formulations are the most popular because they are much simpler and more versatile to study a wide variety of quantum phenomena. The Schrödinger equation in the simple case of a particle with mass m subjected to a conservative potential V(**r**) it is given by [9−13]

$$i\hbar \,\partial\psi(\mathbf{r},t)/\partial t = H\,\psi(\mathbf{r},t) = [-(\hbar^2/2m)\Delta + V(\mathbf{r})]\,\psi(\mathbf{r},t) \qquad (2.1),$$

where H = $-(\hbar^2/2m)\Delta + V(\mathbf{r})$ is the time independent hamiltonian operator, $\Delta$ the laplacian operador and $\psi(\mathbf{r},t)$ is the wavefunction that represent the *state* of the system. It is obtained integrating (2.1) taking into account the *initial state* $\psi(\mathbf{r},t = t_o)$ in a given (initial) time $t_o$ and the boundary condition $\{\psi(\mathbf{r},t)\}_{\mathbf{r}\,\epsilon\,S}$ where S is a surface which involves completely the system.

According to the fundamental postulate[6] of QM all information about the physical state of the system are obtained through the wave function $\psi$(r, t) which is a vector defined in a Hilbert space. Two typical cases[12-14] and illustrative applications in basic courses of (2.1) are: particle in a rectangular box with insurmountable walls and the hydrogen atom. In these and other similar cases, in general, the particles assume discrete values of energy $E_n$ and are described by a discrete spectrum of wavefunctions $\psi_n(\mathbf{r}, t) = \varphi_n(\mathbf{r}) \exp(-iE_n t/\hbar)$ (are the stationary states). The $E_n$ values are determined by the boundary condition $\{\psi(\mathbf{r}, t)\}_{\mathbf{r} \in S}$. As H is a linear operator, the general solution of an arbitrary function $\psi(\mathbf{r},t)$ of (2.1) is given by the superposition

$$\psi(\mathbf{r}, t) = \Sigma_n C_n \varphi_n(\mathbf{r}) \exp(-iE_n t/\hbar) \qquad (2.2)$$

where $C_n$ are arbitrary constants that are determined knowing the wavefunction $\psi(\mathbf{r}, t = t_o)$. Due to the wave nature of particles it makes no sense to calculate its positions $\mathbf{r}(t)$. We can only talk about the probability $dP(\mathbf{r})$ to find a particle in a volume element $d^3\mathbf{r}$ about the point $\mathbf{r}$ defined by $dP(\mathbf{r}) = |\psi(\mathbf{r}, t)|^2 d^3\mathbf{r}$. Thus, the probability distribution of the coordinates $\mathbf{r}$ is given by $|\psi_n(\mathbf{r}, t)|^2 = |\varphi_n(\mathbf{r})|^2$ that is independent of time. With the normalization condition of probability $\int |\psi(\mathbf{r}, t)|^2 d^3\mathbf{r} = 1$ and taking into account the orto–normality of the functions $\varphi_n(\mathbf{r})$ we get $\Sigma_n |C_n|^2 = 1$, where $|C_n|^2 = P_n$ is the probability of finding a state $\varphi_n(\mathbf{r})$ with energy $E_n$. The expected value (average) $<F>$ of an observable F, independent of the time, is given by [12-14]

$$<F> = \int \psi(\mathbf{r}, t) F \psi(\mathbf{r}, t)^* d^3r = \Sigma_n |C_n|^2 F_n \qquad (2.3),$$

where $F_n = \int \varphi_n(\mathbf{r})^* F \varphi_n(\mathbf{r}) d^3\mathbf{r}$. If the system has only a single energy $E_s$ its state is represented by $\psi_s(\mathbf{r},t) = \varphi_s(\mathbf{r}) \exp(-iE_s t/\hbar)$. The spectral lines are generated by transitions between these energy states.

Thus, we see that the particle wave property leads to a probabilistic interpretation for the position $\mathbf{r}$, states $\varphi_n(\mathbf{r})$ and energies $E_n$. This is an intrinsic probabilistic nature of the QM which is a result of (2.1). It is different from the randomness of the experimental results which is generated by measurements made on the microscopic system (see next section). Similar results are found using the quantum relativistic equations.[6,10–12]

2.1) *Causality in Quantum Mechanics.*

The basic problem of quantum dynamics is to calculate the function $\psi(\mathbf{r}, t)$ at any instant t knowing $\psi(\mathbf{r}, t_o)$ a given time $t = t_o$ (initial time). To do this we will show that (2.1) is closely related to the temporal evolution of $\psi(\mathbf{r}, t)$. So, let's assume there is a linear operator $T(t, t_o)$, which is independent of $t_o$, so that[12]

$$\psi(\mathbf{r}, t) = T(t, t_o) \psi(\mathbf{r}, t_o), \qquad (2.4).$$

Therefore, as $\psi(\mathbf{r},t_2) = T(t_2, t_1) \psi(\mathbf{r},t_1) = T(t_2, t_1) T(t_1, t_o) \psi(\mathbf{r},t_o)$ the operator T would obey a *group* property[15] $T(t_2, t_o) = T(t_2, t_1) T(t_1, t_o)$. Consequently, $T(t, t_o) T(t_o, t) = T(t_o, t) T(t, t_o) = 1$ and $[T(t, t_o)]^{-1} = T(t_o, t)$. From the definition (2.4) it is clear that $T(t, t) = 1$. For small $\varepsilon$ we can write, defining an operator $H(t)$:

$$T(t + \varepsilon, t) = 1 - (i/\hbar)\varepsilon H \quad (2.5).$$

Using (2.5) and the group property $T(t + \varepsilon, t_o) = T(t + \varepsilon, t) T(t, t_o)$ we obtain a differential equation for T,

$$dT(t, t_o)/dt = \lim_{\varepsilon \to 0} \{[T(t + \varepsilon, t_o) - T(t, t_o)]/\varepsilon\} = -(i/\hbar)H(t) T(t, t_o),$$

that is,,

$$i\hbar\, dT(t, t_o)/dt = H(t) T(t, t_o) \quad (2.6),$$

with initial condition $T(t_o, t_o) = 1$. From the above we find that $\psi(\mathbf{r}, t + \varepsilon) = T(t + \varepsilon, t) \psi(\mathbf{r},t)$ or, to first order in $\varepsilon$,

$$\psi(\mathbf{r},t) + \varepsilon\, d\psi(\mathbf{r},t)/dt = [1 - (i/\hbar)\varepsilon H(t)] \psi(\mathbf{r},t),$$

consequently,

$$i\hbar\, d\psi(\mathbf{r},t)/dt = H(t) \psi(\mathbf{r},t) \quad (2.7).$$

Which is the general evolution law for the quantum state of a given system. Comparing (2.1) and (2.7) we see that the temporal evolution of $\psi(\mathbf{r}, t)$ is closely related to the Hamiltonian operator H. When H is time independent, applying repeatedly the group property for T in n intervals for $\varepsilon = (t-t_o)/n$ we have, with $T(t_o, t_o) = 1$,

$$T(t, t_o) = \lim_{\varepsilon \to 0, n \to \infty} [1-(i/\hbar)\varepsilon H]^n = \lim_{n \to \infty} [1-(i/\hbar)(t-t_o)H/n]^n,$$

from which we deduce the equation

$$T(t, t_o) = \exp[-(i/\hbar)(t-t_o)H] \quad (2.8),$$

showing that the operator $T(t, t_o)$ is unitary if H is hermitean[10,12–14] and that it is a function of H. We conclude that, at least in the case when H is time independent, there is a time evolution operator which satisfies the equation (2.4), i.e., $\psi(\mathbf{r}, t) = T(t, t_o) \psi(\mathbf{r}, t_o)$. This result defines causality in QM. If H is a perfectly known function of the parameters and physical variables defined in (2.1) the wave function $\psi(\mathbf{r}, t)$ will be mathematically known for any time t (t ≥ $t_o$ and t ≤ $t_o$). Under these conditions it is said that the MQ is *causal* and *deterministic*. In QM (as in CM) the concept of causality applies only if the system evolves undisturbed. The temporal evolution of ψ(**r**, t) = T(t, $t_o$) ψ(**r**, $t_o$) according to (2.8), should take into account only the operator H given by (2.1). It is important to note that *causality* and *determinism* in QM refer to the function ψ(**r**, t), but not to the *physical quantities* (the *observables*) involved in the phenomenon described by equation (2.1). This is completely different from CM where *causality* and *determinism* refers to the *observable* **r**(t) of a macroscopic object. Every time that we submit a quantum system to a measurement process we inevitably provoke a serious perturbation of the system which will totally destroy the causality described by (2.8) which is valid for an isolated system. The measurement act is an external interaction (not contained in H) that we do not know describe. If we are measuring, for example, the position **r** of the particle the observed measured values vary *randomly* within a sphere whose radius goes from zero to infinity. After a large number of measurements we find that the **r** are distributed according to a probability function P(**r**) such that the probability dP(**r**) to find them in a volume element $d^3$**r** about the point **r** is given by dP(**r**) = |ψ(**r**,t)|$^2$ $d^3$**r**, as predicted by (2.2). Other observables can also have random values when measured, but these values will always obey probability functions obtained with a quantum equation like (2.1). Measurements break quantum causality,[6] they introduce a *randomness* in the measured values of observables. It seems there is no functional relationship connecting the random effect of the measurements and quantum causality. This aspect is contained in the Heisenberg *uncertainty relations* [9.12] that are obtained using the Schrödinger equation relating observables that do not commute, such as **r** and **p**, E and t, $L_z$ and φ…Through the quantum dynamics involving the S matrix (defined from (2.8)) and using the perturbation theory [10–12] we can calculate the transition probabilities between quantum states. These and other aspects found in innumerous books and articles written over a century of application of QM show that it is an *incomplete* theory. That is, MQ is unable to determine (predict) the exact values of all observables involved in a given quantum process. The fact that quantum theory is incomplete was asked in 1935 by Einstein, Podolsky and Rosen[16] and from that date until today many articles and book chapters were published analyzing this question.[17]

## 3) Classical and Quantum Correlations.

In 1964 Bell[18] published a theorem that is of fundamental importance for contemporary physics. This is known as Bell's theorem, demonstrated in the form of inequalities. Bell's theorem can be proved based on phenomenological arguments, using common sense, without making use of concepts such as hidden variables (see Section 5) or similar hypotheses.[19]
In this theorem are taken into account correlations between measurements of physical quantities of two parts of a system (isolated from the universe) that at a given moment are very far from each other, but that were interacting in the past. At the moment of the measurements each observer completely ignores what is happening with the other. They are in a "spacelike"[4] condition. Within the context of the SR there is no possible communication between them. So, it is reasonable to expect that the actions of an observer can not influence the results of the measurements made by the other. Let us see how these measurements are made in classical and quantum systems following the Peres article.[19]

(A) *Correlations in Classical Mechanics*.

Consider a bomb that explodes at rest into two asymmetrical parts, each part carrying angular momentum $\mathbf{J}_1$ and $\mathbf{J}_2 = -\mathbf{J}_1$ (see Figure1). We will assume that the measuring devices are located far apart from each other in A and B equidistant from the origin O. Thus the two parts of the bomb will be detected *simultaneously* in A and B. The observer A measures $\mathbf{J}_1$ along a fixed unit vector $\boldsymbol{\alpha}$ and B measures $\mathbf{J}_2$ along a fixed unit vector $\boldsymbol{\beta}$, $\boldsymbol{\alpha}$ and $\boldsymbol{\beta}$ in arbitrary directions. Let us define the parameters

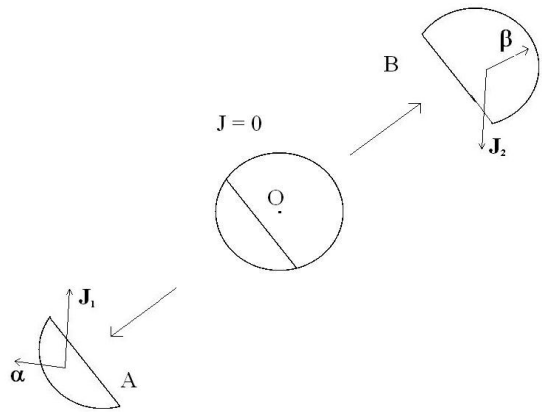

**Figure 1**. A bomb, initially at rest, explodes into two fragments carrying opposite angular moments $\mathbf{J}_1 = -\mathbf{J}_2$. These fragments are detected at the points A and B very far from O.

$r_\alpha$ and $r_\beta$ at A e B, respectively, by

$$r_\alpha = \mathrm{sign}(\boldsymbol{\alpha}\cdot\mathbf{J}_1) \quad \text{and} \quad r_\beta = \mathrm{sign}(\boldsymbol{\beta}\cdot\mathbf{J}_2) \qquad (3.1),$$

that assume the values $r_{\alpha,\beta} = \pm 1$ ou $0$.

Let us suppose that the directions of $\mathbf{J}_1$ and $\mathbf{J}_2$ are unpredictable and randomly distributed. Taking into account that the measurements are repeated N times, their average values are given by

$$< r_\alpha > = (1/N) \Sigma_j\, r_{j\alpha} \quad \text{and} \quad < r_\beta > = (1/N) \Sigma_j\, r_{j\beta},$$

where $r_{j\alpha}$ and $r_{j\beta}$ ( $j = 1,...,N$) denote $r_\alpha$ and $r_\beta$ for the $j^{th}$ explosion. Due to the random distribution of $\mathbf{J}$ it is expected that the only acceptable mean values are practically zero (typically of order $1/\sqrt{N}$). On the other hand, if the observers compare the results after they have been obtained, the average correlation between the measured values $r_\alpha$ and $r_\beta$ at A and B, respectively, which is given by

$$< r_\alpha\, r_\beta > = C_c(\alpha,\beta) = (1/N) \Sigma_j\, r_{j\alpha}\, r_{j\beta} \qquad (3.2),$$

can be non zero. For example, if $\boldsymbol{\alpha} = \boldsymbol{\beta}$, since $r_{j\alpha} = - r_{j\beta}$, we have $< r_\alpha\, r_\beta > = -1$. If $\boldsymbol{\alpha} = -\boldsymbol{\beta}$ we get $< r_\alpha\, r_\beta > = 1$.

Consider a unit sphere cut by a plane perpendicular to $\boldsymbol{\alpha}$. If $\mathbf{J}_1$ points to the upper hemisphere $r_\alpha = 1$ in this hemisphere and $r_\alpha = -1$ at the bottom. Similarly, a second plane perpendicular to the equatorial $\boldsymbol{\beta}$ determines two regions with $r_\beta = \pm 1$. The angle between $\boldsymbol{\alpha}$ and $\boldsymbol{\beta}$ is taken equal to $\theta$. With this procedure the unit sphere is divided into four regions with $r_\alpha r_\beta = \pm 1$. The unit sphere in these four regions delimit areas that are in the ratio of $\theta:(\pi-\theta)$. Assuming that the $\mathbf{J}_1$ and $\mathbf{J}_2$ are randomly distributed, we obtain for large values of N the classical correlation function

$$< r_\alpha\, r_\beta > = C_c(\alpha,\beta) = [\theta - (\pi - \theta)]/\pi = 2\theta/\pi - 1 \qquad (3.3).$$

It is important to emphasize that the hypothesis of randomness of $\mathbf{J}_1$ and $\mathbf{J}_2$ is naturally assumed in classical physics. (B) *Correlations in Quantum Mechanics*.

Let us calculate the correlations using the QM in the case of a particle at rest with zero spin, which disintegrates into two spin ½ fermions in a singlet state $|\Psi>$, as occurs, for example, with $\pi_o$ which decays into an electron and a positron, $\pi^o \rightarrow e^- + e^+$. Being $\mathbf{s}_1$ and $\mathbf{s}_2$ spin operators [10–12] of the particles 1 and 2, respectively, we have $\mathbf{s}_1|\Psi> = -\mathbf{s}_2|\Psi>$. Similarly to the case (A) we define the quantities $r_\alpha = 2\boldsymbol{\alpha}\cdot\mathbf{s}_1$ e $r_\beta = 2\boldsymbol{\beta}\cdot\mathbf{s}_2$ that assume the values $r_{\alpha,\beta} = \pm 1$. The average value $r_\alpha$ is now given by

$$< r_\alpha > = (1/N) <\Psi|\Sigma_j (2\boldsymbol{\alpha}\cdot\mathbf{s}_{1j})|\Psi> = (1/N) <\Psi|\Sigma_j (\boldsymbol{\alpha}\,\boldsymbol{\sigma}_j)|\Psi> =$$
$$= \boldsymbol{\alpha}\, \{(1/N)\Sigma_j <\Psi|\boldsymbol{\sigma}_j|\Psi>\} = 0$$

to a very large number N of measurements, using the **σ** Pauli matrix [10–12] and taking into account that the spin of particles is ½. Similarly we have $<r_\beta> = 0$. The quantum correlation function $C_q(\alpha,\beta) = <r_\alpha r_\beta>$ is now given by

$$< r_\alpha r_\beta > = (1/N) <\Psi | \Sigma_j r_{j\alpha} r_{j\beta} |\Psi > = (1/N)\{\Sigma_j <\Psi |(2\boldsymbol{\alpha}\cdot\mathbf{s}_{1j})( 2\boldsymbol{\beta}\cdot\mathbf{s}_{2j})|\Psi >\} =$$

$$= (1/N) \{\Sigma_j <\Psi |(\boldsymbol{\alpha}\cdot\boldsymbol{\sigma}_{1j})(\boldsymbol{\beta}\cdot\boldsymbol{\sigma}_{2j})|\Psi >\} =$$

$$= (1/N) \{\Sigma_j <\Psi |[(\boldsymbol{\alpha}\cdot\boldsymbol{\beta})-i\boldsymbol{\sigma}_j\cdot(\boldsymbol{\alpha} \times \boldsymbol{\beta})|\Psi >\}$$

$$= \boldsymbol{\alpha}\cdot\boldsymbol{\beta} -i (\boldsymbol{\alpha} \times \boldsymbol{\beta})\cdot\{(1/N)\Sigma_j <\Psi | \boldsymbol{\sigma}_j |\Psi >\} = \boldsymbol{\alpha}\cdot\boldsymbol{\beta} + 0,$$

where we have put $\boldsymbol{\sigma}_j = \boldsymbol{\sigma}_{1j} = -\boldsymbol{\sigma}_{2j}$ and taken into account the identity[10] $(\boldsymbol{\alpha}\cdot\boldsymbol{\sigma}_{1j})(\boldsymbol{\beta}\cdot\boldsymbol{\sigma}_{2j}) = -\boldsymbol{\alpha}\cdot\boldsymbol{\beta} - i\boldsymbol{\sigma}_j\cdot(\boldsymbol{\alpha} \times \boldsymbol{\beta})$.

So, the quantum correlation function $C_q(\alpha,\beta)$ becomes

$$< r_\alpha r_\beta > = C_q(\alpha,\beta) = -\boldsymbol{\alpha}\cdot\boldsymbol{\beta} = -\cos\theta \qquad (3.4),$$

where θ is the angle between the **α** and **β**. Comparing classical $C_c(\alpha, \beta) = 2\theta/\pi -1$ and quantum $C_q(\alpha, \beta) = -\cos\theta$ correlation functions we found that, as a function of θ, $|C_q(\alpha, \beta)| \geq |C_c(\alpha, \beta)|$.

### 4) Bell's Theorem.

As said before, if observers A and B are too far apart, at a "spacelike" condition,[4] within the context of SR there is no possibility of communication between them. In this way it is reasonable that the actions of an observer can not influence the results of the measurements made by the other. Thus, we expect that the results obtained by A are independent of what B is doing and vice versa. If, for example, B has oriented its apparatus of measurement along a different direction **β'**, the results obtained by A should be exactly the same, not only $<r_\alpha> = 0$, but also each individual value $r_{\alpha j}$ (j = 1 ,..., N) should remain unchanged. This should happen even if the bomb fragments or particles with spin ½ have begun its flight before A and B orientate their apparatus along **α** and **β.** This hypothesis is named *locality hypothesis* that in a shorthand way means that the message detected in B depends only on the changes made at B, not on the position of the detector A and vice versa.

Bell's theorem[17,18] is demonstrated in the form of inequalities. We will show here the theorem for a single type of inequality following the article Peres.[19] For measurements effectively carried out, even with a new orientation **α**´and **β**´ we have only $r_{α'} = ± 1$ e $r_{β'} = ± 1$ and, furthermore, acceptable measurements must obey the conditions $<r_{α'}> = 0$ and $<r_{β'}> = 0$. In this way it is easy to verify by simple inspection, that for any possible choices of $r_α$, $r_β$, $r_{α'}$ e $r_{β'}$ we must have

$$r_α r_β + r_α r_{β'} + r_{α'} r_β - r_{α'} r_{β'} = ± 2 \qquad (4.1).$$

Consequently,

$$(1/N)| \Sigma(r_α r_β + r_α r_{β'} + r_{α'} r_β - r_{α'} r_{β'})| ≤ 2 \qquad (4.2).$$

Defining $C(p,q) = <r_p r_q>$ we see that (4.2) becomes written as

$$|C(α,β) + C(α,β´) + C(·α´,β) - C(α´,β´)| ≤ 2 \qquad (4.3),$$

that is one of the inequalities which constitute the famous Bell´s theorem. It is very important to note that inequality (4.3) was derived without making any assumptions about the physical theory involved in the correlation measurements. It is precisely here that lies the great power of this theorem: the demonstration does not depend on a particular physical theory. Thus, if (4.3) is violated in quantum phenomena experiments it means that this violation is an intrinsic property of the quantum phenomena. It is not due to a given quantum theory used to explain the quantum effects. It can be the QM[10−12] or a *hidden variable theory* (see Section 5). If the inequality (4.3) is experimentally violated would imply that *quantum phenomena are non−local.*

It is easy to show that (4.3) is violated *theoretically* in the QM framework. Indeed, it is enough to use the function $C_q(α, β) = -\mathbf{α·β}$ considering, for example, the case when **α**, **β**, **α**´and **β**´ are coplanar with **α = β, with α´and β´** at an angle ζ one on each side of **α** and **β**. In this case **α·β** = 1, **α·β**´ = **α´·β** = cosζ and **α´·β´** = cos2ζ. So, taking this into account (4.3) we see that

$$|-1 - 2 + \cosζ \cos2ζ| = f(ζ) \qquad (4.4).$$

In Figure 2 where is shown f(ζ) as a function of ζ (measured in degrees) we see that Bell's inequality (4.3) is violated, that is, f(ζ) > 2 for angles ζ < 90° (values above the dashed line − − −).

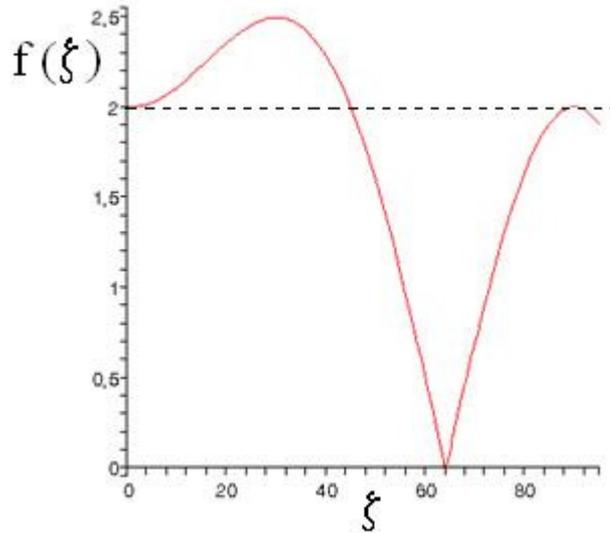

**Figure 2**. f(ζ) as a function of the angle ζ , measured in degrees. We see that f(ζ) > 2 for angles ζ < 90º (values above the dashed curve).

Theoretical violations of the (4.3) in QM do not occur only for spins and photons correlations which have the same algebra. These violations are present for any pair of states of any two suitably correlated systems.[19] The correlation effect between these states is known as *entanglement effect in quantum mechanics.*

Finally, let us see what happens *theoretically* in CM taking into account that $C_c(α,β) = 2θ/π − 1$, according to (3.3), where θ is the angle between **α** e **β.** For the same conditions adopted above for **α, β, α´** e **β´** we verify that (4.3) becomes

$$| −1 + 2(−1 + 2ζ/π) − (−1 + 4ζ/π ) | = 2,$$

for any values ζ, obeying the Bell´s inequality (4.3)

Repeating this kind of analysis for all possible orientations of **α, β, α´** and **β'** we can verify that, using the *QM formalism*, there will be always situations in which inequality (4.3) will be violated. Unlike, in *CM* inequality (4.3) will always be obeyed, that is, classical correlations do not violate Bell's inequality.

Very sophisticated and precise experiments show,[17,20] with very good approximation, that Bell's inequality (4.3) is violated by measuring correlations in quantum phenomena.

4.1) *Conclusions.*

Considering that very sophisticated experiments[17, 20] confirm very precisely that Bell's inequality (4.3) is violated in quantum phenomena we deduce that there must be transmission of an instantaneous signal between points A and B that are responsible for the correlations between the $r_α$ and $r_\beta$ measurements. Thus, we conclude that *quantum phenomena are non−local*. Note that this result, as comments done in Section 3, is independent of the quantum theory: it can be the QM[10-12] or a hidden variable theory (see Section 5). Thus, taking into account these results and also what was analyzed in Section 2 we can conclude that QM is *incomplete and non−local theory*.

It remains the unsolved problem of supra−luminous quantum signal that seems to violate the SR. This paradox give rise to a new branch of physics called quantum cryptography.[21]

## 5) Hidden Variables Theory and QM. Final Conclusions.

We will briefly review the meaning of *hidden variables theory*[22] that was proposed due to the statistical nature of QM. The non −completeness of the QM left and still leaves many physicists dissatisfied. To these it would be necessary to construct a new theory adopting different variables from those used in QM in order to get a precise description of the physical reality. This new theory is known as *hidden variables theory*. Anyway, according to the experimental results arising from the violation of Bell's theorem, a *hidden variables theory* or QM are *non−local theories*. This would invalidate, for example, the Einstein, Podolsky and Rosen proposition[22] based on the EPR paradox.

In one century of existence it was found a successful agreement between QM predictions and the experimental results for almost all quantum phenomena. Due to this and also due to the amazing experimental results of the Bell's theorem investigations most physicists agree that the current MQ and the essence of nature are beyond the limits of classical physics and relativity. That majority believes that the true theory of the universe is the QM, in spite of being non−local and incomplete. The QM alone shows no internal inconsistency and in the macroscopic limit gives the CM. However, hopes for a local theory of hidden variables are still very much alive.

A hidden variables theory which has a fair popularity among physicists was proposed by Bohm in 1982.[23] It was built based on the Hamilton−Jacobi theory[7] where the movement of a particle is governed by a "wave guide". In this context the "position" and "momentum" of a particle defined in a configuration space are the *hidden variables*. As their predictions are identical to those obtained by QM[23,24] it is also an incomplete theory.